\documentclass[prb,twocolumn,showpacs,floatfix,preprintnumbers,amsmath,amssymb,superscriptaddress]{revtex4}
\usepackage{graphicx}
\usepackage{dcolumn}
\usepackage{bm}
\usepackage{color}
\usepackage{color}
\usepackage[latin1]{inputenc}
\usepackage[urlcolor=blue]{hyperref}
\hypersetup{backref, colorlinks=true} 

\begin{document}

\preprint{APS/123-QED}

\title{{Temperature dependence of the lower critical field $H_{c1}(T)$ evidences nodeless superconductivity in FeSe}}

\author{M. Abdel-Hafiez}\email{m.mohamed@ulg.ac.be}
\affiliation{D\'epartement de Physique, Universit\'e de Li\`ege, B-4000 Sart Tilman, Belgium.}

\author{J. Ge}
\affiliation{INPAC -- Institute for Nanoscale Physics and Chemistry, Nanoscale Superconductivity and Magnetism Group, K.U.Leuven, Celestijnenlaan 200D, B--3001 Leuven, Belgium.}

\author{A. N. Vasiliev}
\affiliation{Low Temperature Physics and Superconductivity Department, M.V. Lomonosov Moscow State University, 119991 Moscow, Russia.} \affiliation{Theoretical Physics and Applied Mathematics Department, Ural Federal University, 620002 Ekaterinburg, Russia.}

\author{D. A. Chareev}
\affiliation{Institute of Experimental Mineralogy, Russian Academy of Sciences, 142432 Chernogolovka, Moscow District, Russia.}

\author{J. Van de Vondel}
\affiliation{INPAC -- Institute for Nanoscale Physics and Chemistry, Nanoscale Superconductivity and Magnetism Group, K.U.Leuven, Celestijnenlaan 200D, B--3001 Leuven, Belgium.}

\author{V. V. Moshchalkov}
\affiliation{INPAC -- Institute for Nanoscale Physics and Chemistry, Nanoscale Superconductivity and Magnetism Group, K.U.Leuven, Celestijnenlaan 200D, B--3001 Leuven, Belgium.}

\author{A. V. Silhanek}
\affiliation{D\'epartement de Physique, Universit\'e de Li\`ege, B-4000 Sart Tilman, Belgium.}

\date{\today}

\begin{abstract}
We investigate the temperature dependence of the lower critical field $H_{c1}(T)$ of a high-quality FeSe single crystal under static magnetic fields $H$ parallel to the $c$ axis. The temperature dependence of the first vortex penetration field has been experimentally obtained by two independent methods and the corresponding $H_{c1}(T)$ was deduced by taking into account demagnetization factors. A pronounced change of the $H_{c1}$(T) curvature is observed, which is attributed to multiband superconductivity. The London penetration depth $\lambda _{ab}(T)$ calculated from the lower critical field does not follow an exponential behavior at low temperatures, as it would be expected for a fully gapped clean $s$-wave superconductor. {Using either a two-band model with $s$-wave-like gaps of magnitudes $\Delta _{1}$ = 0.41 $\pm$ 0.1\,meV and $\Delta _{2}$ = 3.33 $\pm$ 0.25\,meV or a single anisotropic s-wave order parameter, the temperature-dependence of the lower critical field $H_{c1}(T)$ can be well described. These observations clearly show that the superconducting energy gap in FeSe is nodeless.}
\end{abstract}

\pacs{74.20.Rp, 74.25Ha, 74.25.Dw, 74.25.Jb, 74.70.Dd}
\maketitle

\section{Introduction}
{Superconductivity in Fe-based superconductors (FeSCs) has been studied intensively due to the relatively large transition temperatures $T_{c}$ up to 55~K,~\cite{ZA,HH,JY,CW} a high upper critical field, and a layered structure similar to the cuprates. } Superconductivity emerges when the magnetic order is suppressed by charge doping or by applying an external pressure, and takes place within FeAs, FeP, or FeSe crystallographic planes. These planes are separated by layers of other elements serving as a charge reservoirs. Among the very few members of this FeSCs class becoming superconducting at ambient pressure and without doping, we find the iron selenide, FeSe. This compound has the simplest crystal structure and stoichiometry, while keeping a moderate superconducting critical temperature of about 8\,K for polycrystalline samples.~\cite{YM} Furthermore, high-quality single crystals with rather large dimensions can be grown,~\cite{AE,DC,Hu} which are necessary for an accurate determination of {bulk} physical properties. Interestingly, if the tetragonal FeSe system is submitted to a pressure of 8.9\,GPa, a huge enhancement of the $T_{c}$ up to 36.7\,K is obtained.~\cite{Med,YM} This system also stands out due to the absence of nesting between hole and electron pockets of the Fermi surface.~\cite{YM} {In addition, density functional calculations of the electronic structure indicate that the electron-phonon coupling cannot explain superconductivity at such a high transition temperature.~\cite{Subedi}} Therefore, FeSe falls in the category of unconventional superconductivity and appears as an ideal candidate to study the fundamental properties of superconductivity in clean iron based superconductors.


One of the crucial issues to elucidate the mechanism leading to high-temperature superconductivity is the nature of pairing, e.g., the symmetry and structure of the superconducting order parameter. Up to now, there have been several investigations on the pairing symmetry of FeSe superconductors. Thermal conductivity measurements~\cite{JKD} show the absence of nodes in the superconducting gap. Furthermore, recent upper critical field studies of $\beta$- FeSe crystals have revealed that two-band effects dominate $H_{c2}(T)$, with possible influence of a spin paramagnetic effect.~\cite{Hu2} The presence of both an isotropic $s$-wave and extended $s$-wave order parameters coexisting in a superconducting single-crystal FeSe has also been proposed based on specific-heat measurements.~\cite{JYLin} Very recently, multiple Andreev reflections spectroscopy pointed to the existence of two-gap superconductivity.~\cite{DC} In addition, muon spin rotation studies of the penetration depth $\lambda _{ab}^{-2}(T)$ in FeSe$_{0.85}$ were consistent with either two-gap $(s+s)$ or anisotropic $s$-wave order parameter symmetries, thus implying that the superconducting energy gap contains no nodes.~\cite{RKH} In contrast to that, scanning tunneling spectroscopy experiments in the stoichiometric FeSe provided clear evidence for nodal superconductivity.~\cite{CLS} The observed gap function was attributed to an extended $s$-wave pairing structure with the mixture of $s_{x^{2}+y^{2}}$ and $s_{x^{2}y^{2}}$ pairing symmetries.~\cite{HHH} Clearly, today there is no general consensus on the origin of superconducting pairing mechanism in FeSe compounds and further measurements to elucidate this issue are necessary.

The lower critical field, $H_{c1}(T)$ i.e. the thermodynamic field at which the presence of vortices into the sample becomes energetically favorable, and the magnetic penetration depth, $\lambda(T)$, are very useful parameters providing key information regarding bulk thermodynamic properties. Indeed, the gap properties of different families of FeSCs have been investigated by tracking the $H_{c1}$(T) and the magnetic penetration depth.~\cite{CR,KSS,YS,XLWang,REN,Martin,RG}The gap properties of these compounds display single to double gaps and even the presence of nodes. This variety of gap properties appears to be related to the nature and the level of doping. {It should be mentioned that this quest for new multiband superconductivity is a timely subject due to the possible emergence of non-monotonic vortex-vortex interactions, fingerprint of so-called type 1.5 superconductivity.~\cite{VVM,Milo,BA}}


{The main motivation of the present work is to tackle a long standing question concerning whether the superconducting properties of these materials can be accounted for a nodal order parameter or not. To that end, we reliably determined the temperature dependence of the $H_{c1}$ from magnetization measurements. We then compare the most popular approach of determining the first vortex penetration as the point of deviation from a linear $M(H)$ response, to the value obtained from the onset of the trapped magnetic moment ($M_{t}$). Although this latter approach to determine $H_{c1}(T)$ has been performed in detail in high-$T_{c}$ superconductors, i.e., YBa$_{2}$Cu$_{3}$O$_{0.69}$ as reported in Ref.~\onlinecite{Moshchalkov}, its application to Fe-based superconductors has not been presented so far. In particular, our results show that the method of the trapped magnetization onset is more sensitive than the method determined from the point of deviation from a linear $M(H)$ response. In addition, a kink around 7\,K is obtained on the $H_{c1}(T)$ curve, which can be accounted for by the multi-band nature of superconductivity in this system. Our analysis further shows that the superconducting gaps determined through fittings to the {in-plane} London penetration depth cannot be described with the single-band weak-coupling BCS scheme; rather, it implies the presence of either two $s$-wave-like gaps with different magnitudes and contributions or a single anisotropic s-wave gap. Our London penetration depth results are contrasted to values obtained through the more sophisticated technique of muon-spin rotation.}~\cite{RKH}


\section{Experimental}
\begin{figure}
\includegraphics[width=20pc,clip]{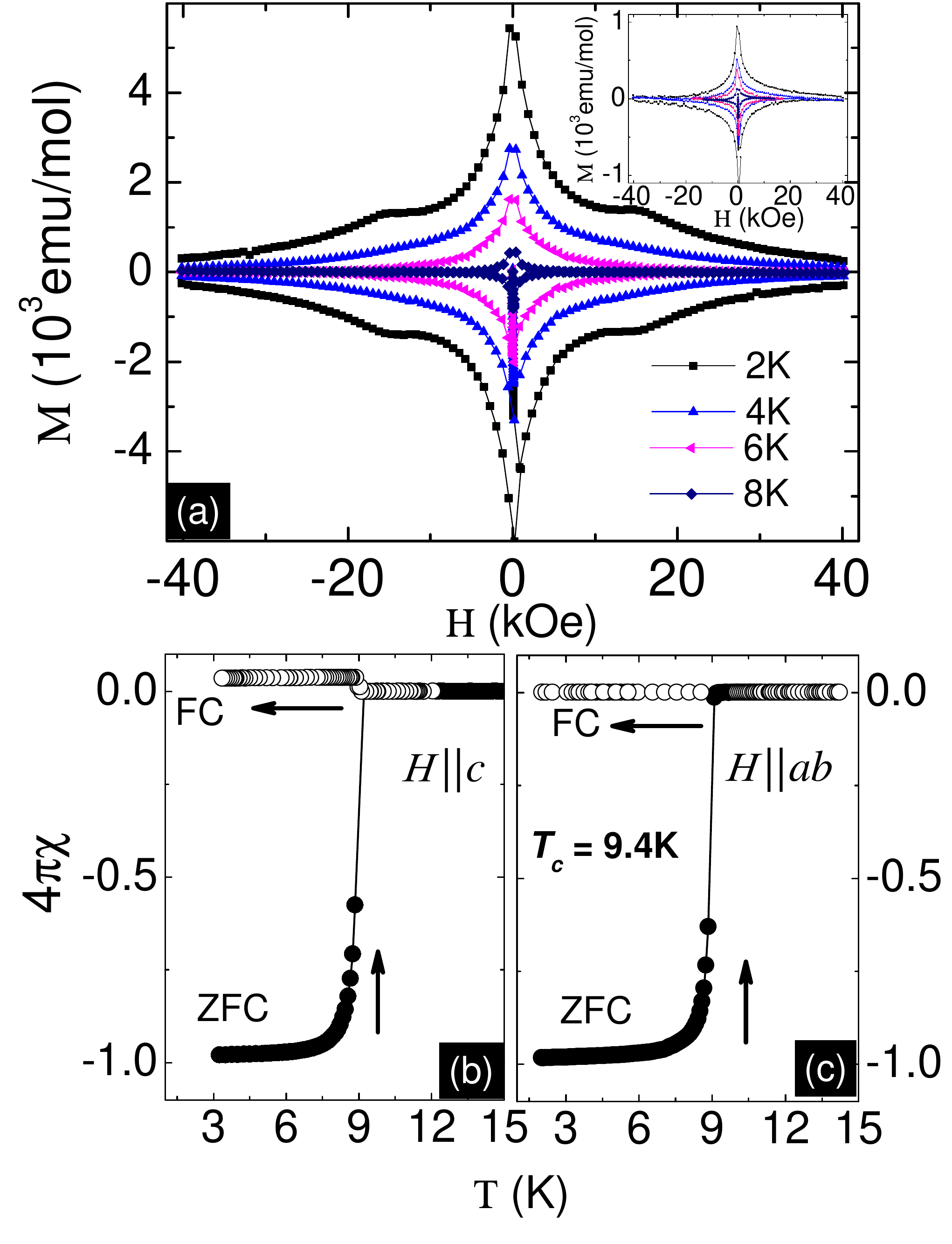}
\caption{(a) Magnetic field dependence of the isothermal magnetization $M$ vs. $H$ loops measured at different temperatures ranging from 2 to 8\,K up to 40\,kOe with the field parallel to both $c$ axis and $ab$ plane as an inset. (b) and (c) present the temperature dependence of the magnetic susceptibility $\chi$ after demagnetization correction in an external field of 1\,Oe applied along $c$ and $ab$, respectively. The $\chi$ has been deduced from the dc magnetization measured with $H \parallel c$  and $H \parallel ab$ following ZFC and FC protocols for FeSe single crystals. } \label{Fig:1}
\end{figure}

We investigated a selected plate-like FeSe single crystal grown in evacuated quartz ampoule using the AlCl$_{3}$/KCl flux technique with a constant temperature gradient of 5\,$^{o}$C/cm along the ampoule length (temperature of the hot end was kept at 427\,$^{o}$C, temperature of the cold end was about 380\,$^{o}$C). The phase purity of the resulting crystal was checked with X-ray diffraction.~\cite{DC} The sample has lateral dimensions $a \times$$b \times$$c$ = 1.05$\pm0.08 \times$1.25$\pm0.1 \times$0.02$\pm0.01$\,mm$^{3}$ with a mass of 1.2\,mg. Magnetization measurements were performed using a superconducting quantum interference device magnetometer (MPMS-XL5) from Quantum Design. The good quality of the crystals is confirmed from a sharp specific-heat jump (9.45 mJ/mol K$^{2}$ with zero residual specific-heat $\gamma _{r}$)~\cite{JYLin} indicative of a complete superconducting volume. The temperature dependence of resistance $R(T)$ demonstrates a metallic behavior with $T_{c}$ = 9.4\,K.~\cite{DC}

\section{Results and discussions}

\subsection{Irreversible magnetization}

Fig.\,1(a) presents the field dependence of the isothermal magnetization $M$ at certain selected temperatures up to 40\,kOe for $H \parallel c$ (main panel) and $H \parallel ab$ (see the inset of Fig.\,1(a)). For $H \parallel c$, the magnetic irreversibility presents a second peak. Whereas no second peak is observed for $H \parallel ab$. This significant anisotropic behavior in the appearance of the second peak has been reported previously in other Fe-based superconductors, e.g., Ba(Fe$_{0.93}$Co$_{0.07}$)$_{2}$As$_{2}$~\cite{PRO} and LiFeAs~\cite{Ashim} single crystals and is typically associated with the nature of pinning.~\cite{DG} As reported in the 122 and 111 systems the second peak is regarded as the crossover from plastic to elastic pinning.~\cite{PRO,Ashim}


Fig.\,1(b) and Fig.\,1(c)  show the temperature dependence of the magnetic susceptibility of the FeSe single crystal measured by following zero-field cooled (ZFC) and field-cooled (FC) procedures in an external field of 1\,Oe applied along $c$ and $ab$ axis, respectively. The ZFC data for both orientations show a sharp diamagnetic signal, thus confirming bulk superconductivity in our investigated system. The magnetic susceptibility exhibits a superconducting transition with an onset transition temperature $T_\mathrm{c}^{\chi}$ of 9.4\,K for both orientations. The clear irreversibility between FC and ZFC measurements is consequence of a strong vortex trapping mechanism, either by surface barriers or bulk pinning.  Notice that the magnetic moment in field cooled conditions for $H \parallel c$ axis becomes positive for $T<T_c$ (see Fig.\,1(b)). A similar {behavior} has been observed in {conventional as well as} Fe-based superconductors.~\cite{Pramanik2011}

The fact that the hysteresis loops for both orientations are symmetric around $M=0$, points to the relatively weak surface barriers and is indicative of strong bulk pinning. This consideration holds for all studied temperatures, even close to $T_c$ and guarantees that vortex penetration occurs at a field close to $H_{c1}$. In contrast to that, if surface barriers were predominant, the first vortex entrance could take place at much higher field ($\sim H_c$). This is a very important point in order to obtain reliable estimations of the thermodynamic lower critical field as we will discuss below. It is worth noting that the superconducting $M(H)$ exhibits a very weak magnetic background. This indicates that the sample contains negligible {amounts of} magnetic impurities.

\begin{figure}
\includegraphics[width=20pc,clip]{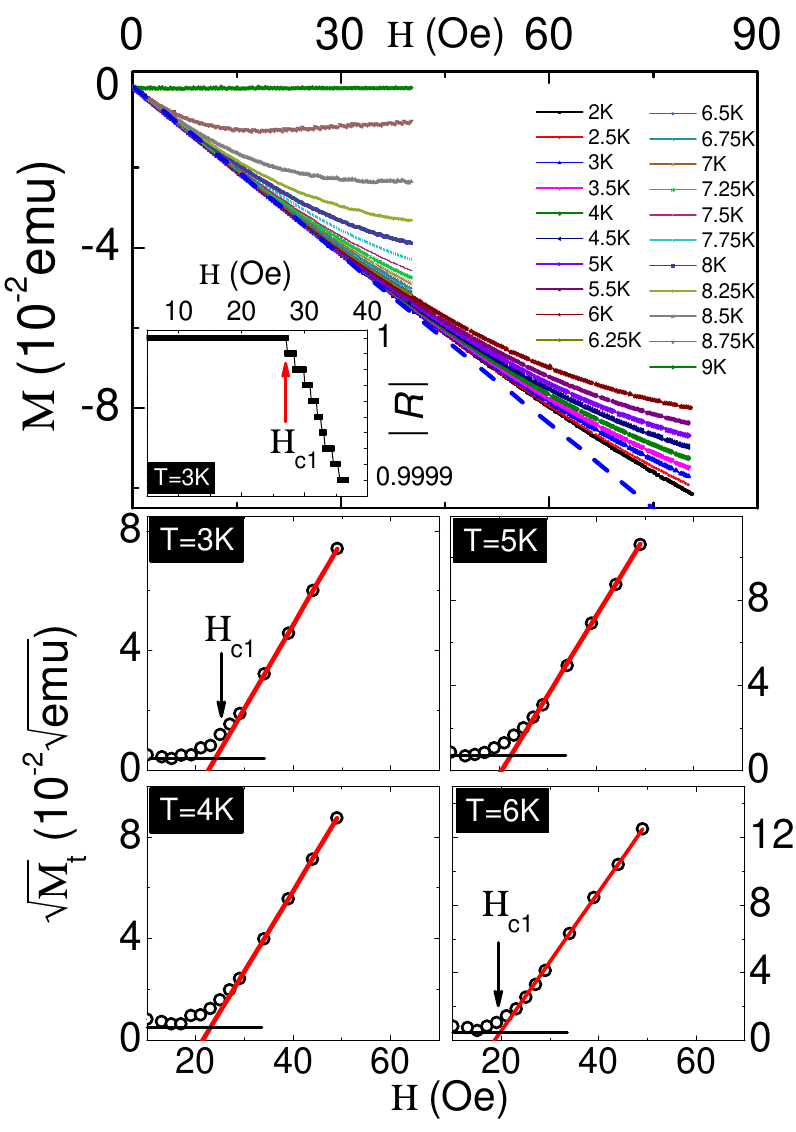}
\caption{The upper panel shows the superconducting initial part of the magnetization curves measured of $\beta$-FeSe single crystals at various temperatures for $H \parallel c$. The dashed line depicts the  Meissner line (linear fits between 0 and 15\,Oe). The inset depicts an example used to determine the $H _{c1}$ value using the regression factor, $R$, at $T$ = 3\,K (see text). The lower panels present the field dependence of the typical plot of $\sqrt{M_{t}}$ vs $H$ at various temperatures. The solid lines are a linear fit to the high-field data of $\sqrt{M_{t}}$ vs. $H$. $H_{c1}$ values are determined by extrapolating the linear fit to $\sqrt{M_{t}}$ = 0.} \label{Fig:1}
\end{figure}

From the magnetization hysteresis loops $M(H)$, we calculated the critical current density $J_{c}$ by using the critical state model with the assumption of field-independent $J_{c}$.~\cite{BB,SH}
\begin{equation}
\label{eq2} J_{c} = \frac{20\Delta M}{[a(1-\frac{a}{3b})]},
\end{equation}
where $\Delta M$ = $M_{dn}-M_{up}$, $M_{dn}$ and $M_{up}$ are the magnetization measured with decreasing and increasing applied field, respectively, $a$ [cm] and $b$ [cm] are sample widths ($a < b$). The unit of $\Delta M$ is in electromagnetic unit per cubic centimeter and the calculated $J_{c}$ is in Ampere per square centimeter. We obtain $J_{c}(2 K) \sim$ 1.34 $\cdot$ 10$^{4}$ A/cm$^{2}$ for $H \parallel c$ and $J_{c}(2 K) \sim$ 1.8 $\cdot$ 10$^{4}$ A/cm$^{2}$ for $H \parallel ab$. These values are lower than those reported in Ba-122, 1111, 11, and the 111 systems ~\cite{BShen,Bhoi,RWHu,Ashim} and higher than those observed in K$_{0.64}$Fe$_{1.44}$Se$_{2}$.~\cite{Lei3}

\subsection{Experimental determination of the lower critical field}

Determining the lower critical field from magnetization measurements has never been an easy task, particularly since $H_{c1}$ is an equilibrium thermodynamic field, whereas the magnetization curve is highly irreversible as a consequence of metastable vortex states far from equilibrium. The most popular method to estimate $H_{c1}$ (here tagged as method-A) consists of detecting the transition from a Meissner-like linear $M(H)$ regime to a non-linear $M(H)$ response, once the vortices penetrate into the sample and build up a critical state. This transition is not abrupt therefore bearing a substantial error bar.

These sort of measurements are obtained by tracking the virgin $M(H)$ curve at low fields at several temperatures, as shown in the upper panel of Fig.\,2 for $H \parallel c$. We have adopted a rigorous procedure (i.e. with user-independent outcome) to determine the transition from linear to non-linear $M(H)$, which consist of calculating the regression coefficient $R$ of a linear fit to the data points collected between $0$ and $H$, as a function of $H$. Then, $H_{c1}$ is taken as the point where the function $R(H)$ departs from 1. This procedure is illustrated for a particular temperature $T = 3$\,K in the inset of the upper panel of Fig.\,2.

An alternative and seemingly more reliable way to determine the lower critical field (here tagged as method-$B$) can be obtained by measuring the onset of the trapped moment $M_{t}$ as described in Refs.~\cite{Moshchalkov,Angst} In contrast to method-$A$ where a heavy data post-processing is needed, now a careful measurement protocol needs to be followed with little data analysis. Indeed, the trapped flux moment $M_{t}$ is obtained by (i) warming the sample up to temperatures above $T_{c}$, ($T$ = 20\,K), then (ii) cooling the sample at zero field down to the chosen temperature, subsequently (iii) the external magnetic field is increased to a certain maximum value $H_{m}$ and (iv) measure the remanent magnetization $M_{t}$ after the applied field has been switched off. The field $H_{m}$ at which $M_{t}$ deviates from zero determines the $H_{c1}$ value at the desired temperature. It is important to notice that this method furnish us with a {rather} independent determination of $H_{c1}$, {weakly} linked to the first procedure (method-$A$) described above.

When taking into account the reversible magnetization, the trapped magnetic moment is $M _{t} \propto (H- H_{c1})^{2}$.~\cite{Moshchalkov} Then the extrapolation $\sqrt{M_{t}}\rightarrow 0$ determines the exact value of the $H _{c1}$. The lower panel of Fig.\,2 presents the typical plot of $\sqrt{M_{t}}$ vs. the applied field, $H$, for our FeSe single crystal. The solid line is a linear fit to the high-field data of $\sqrt{M_{t}}$ vs. $H$. $H_{c1}$ is determined by extrapolating the linear fit to $\sqrt{M_{t}}$ = 0.


Once the values of $H_{c1}$ have been experimentally determined, we need to correct them accounting for the demagnetization effects. Indeed, the deflection of field lines around the sample leads to a more pronounced Meissner slope given by $M/H_{a} = -1/(1-N)$, where $N$ is the demagnetization factor. Taking into account these effects, the absolute value of $H _{c1}$ can be estimated by using the relation proposed by Brandt:~\cite{Brandt}

\begin{equation}
\label{eq2} q_{disk} = \frac{4}{3\pi}+\frac{2}{3\pi}{\tanh[1.27\frac{c}{a}\ln(1+\frac{a}{c})]},
\end{equation}
where $q_{disk} \equiv (|M/H_{a}|-1)(c/a)$, $a$ and $c$ are the dimension perpendicular to the field and thickness of our investigated sample, respectively. For our sample we find $N$ $\approx0.9623$. In addition, an alternative way to determine the demagnetization factor is from rectangular prisms approximation based on the dimensions of the crystal giving us $N$ $\approx0.9688$.~\cite{Pardo}

\begin{figure}
\includegraphics[width=21pc,clip]{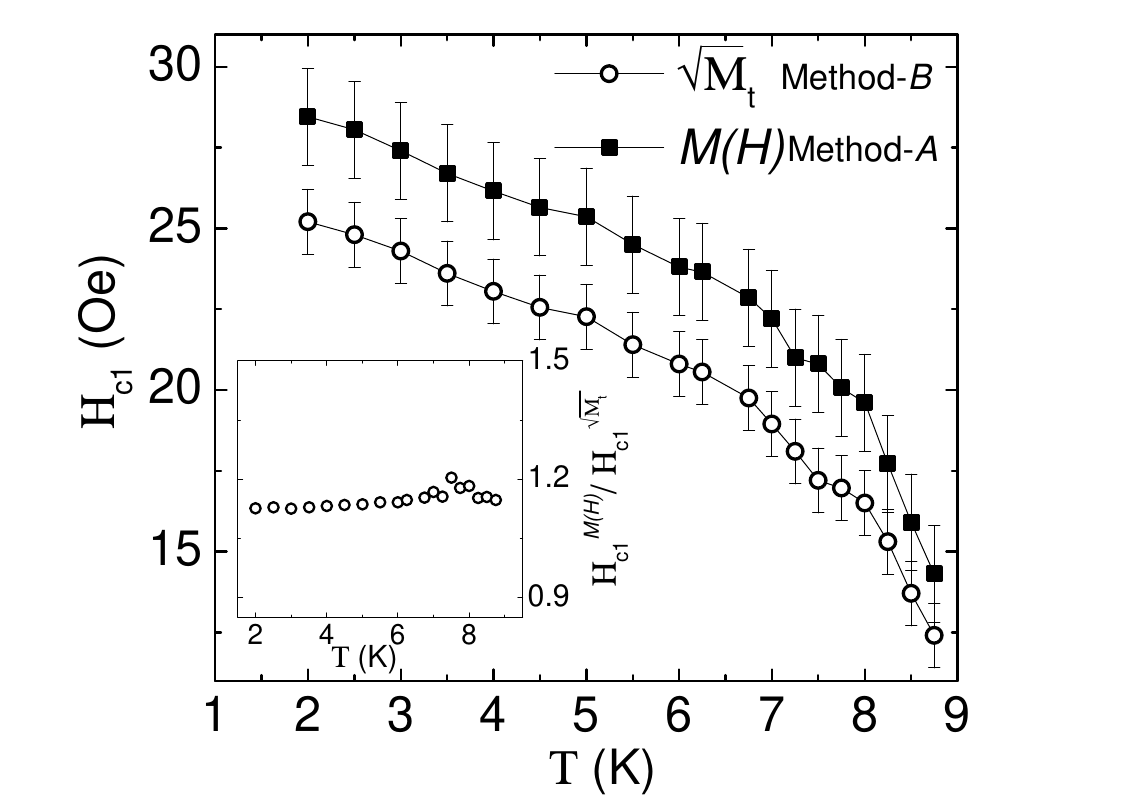}
\caption{The lower critical fields of FeSe single crystal for the field applied parallel to $c$ axis. $H_\mathrm{c1}$ has been estimated by two different methods - from the extrapolation of $\sqrt{M_{t}}\rightarrow 0$ (see the lower panels of Fig.\,2) and from the regression factor (see the inset of the upper panel in Fig.\,2). The bars show the uncertainty of estimated by the deviating point of the regression fits and the linear fit of $\sqrt{M_{t}}$. The inset shows the ratio of $H_{c1}(T)$ values obtained by both methods.} \label{Fig:1}
\end{figure}

The corrected values of $H _{c1}$ obtained by following the two methods described above, are illustrated in Fig.\,3 for $H \parallel c$. Even though both procedures yield different values of $H _{c1}$, the ratio of both methods is just a constant factor with no change on the shape or the dependence (see the inset of Fig.\,3). This fact shows that both methods can provide a qualitative estimation of the temperature dependence of $H_{c1}$. However, method-$B$ shows lower $H _{c1}$ values than method-$A$, which means that the former method is much more sensitive than the latter method. Although to obtain a more quantitative result, we should compare these methods to high resolution imaging techniques such as Magneto-Optical Imaging, Bitter decoration or Scanning Hall Probe Microscopy.

\subsection{Theoretical fitting of the lower critical field}

Irrespective of the used method to obtain $H_{c1}$, a pronounced change of the curvature is observed around 7\,K. This may be attributed to the multi-band nature of superconductivity in our system. This behavior is reminiscent of that reported for the two band superconductors MgB$_{2}$~\cite{Papagelis} and Fe-based superconductors~\cite{CR,YS}, in which similar $H _{c1}(T)$ curves were well fitted by a two-gap weak-coupling BCS model.

Alternatively, in order to shed light on the pairing symmetry in our system, we determined the temperature dependence of the magnetic London penetration depth ($\lambda _{ab}$) applied along the $c$ axis by using the following formula (taking the demagnetization effect into account): $\mu_{0}H_{c1}^{\parallel c} = (\phi$$_{0}/4\pi\lambda _{ab}^{2})\ln\kappa _{c}$, where $\phi$$_{0}$ is the magnetic-flux quantum and $\phi$$_{0}$ = $h/e^{\ast}$ = 2.07 x 10$^{-7}$Oe cm$^{2}$, $\kappa _{c}$ =$\lambda _{ab}$/$\xi _{ab}$ = 72.3 (Ref.~\onlinecite{Lei}) is the Ginzburg-Landau parameter. The results of this calculations are shown in the inset of Fig.\,4. Since we believe that method-$B$ to determine $H _{c1}$ is more accurate, we have calculated $\lambda _{ab}$ only for this method. The penetration depth of our FeSe shows similar behavior to the penetration depths as reported in LiFeAs.~\cite{YS} At low temperatures, $\lambda _{ab}(T)$ does not show the typical exponential behaviour expected for a fully gapped clean $s$-wave superconductor.


Furthermore, if we compare our data to the single-gap BCS theory (i.e., a weak-coupling approach), we find that a single BCS gap cannot be reconciled with our experimental data (see the blue dashed line in Fig.\,4). Indeed, the single BCS gap leads to a rather different trend and shows a systematic deviations from the data in the whole temperature range below $T_\mathrm{c}$. In addition, it largely misses the kink around 7\,K.

\begin{figure}
\includegraphics[width=20pc,clip]{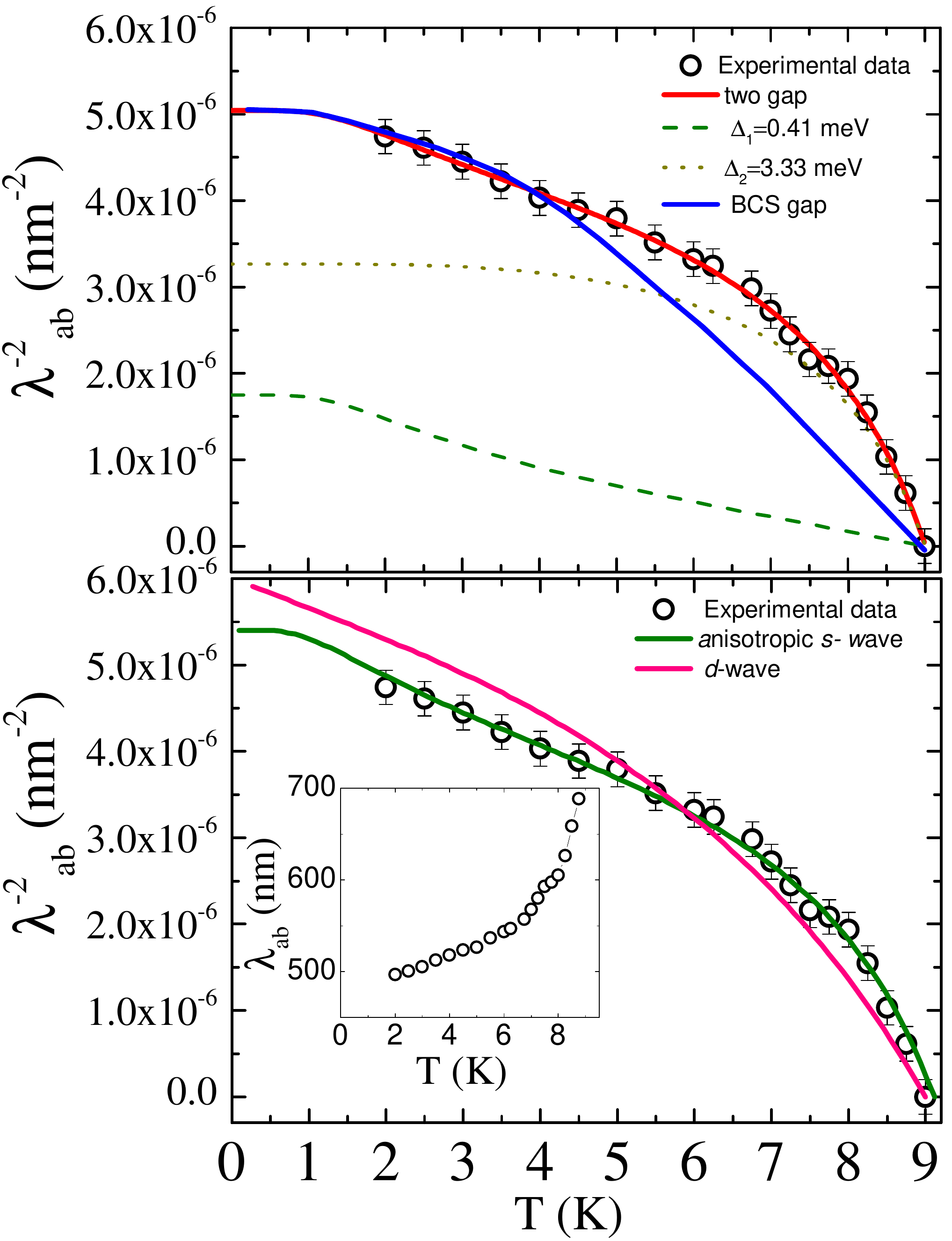}
\caption{{Upper panel: The temperature dependence of the London penetration depth, $\lambda _{ab}^{-2}(T)$, for FeSe. The solid red line is the fitting curve using the two-gap model. The dashed and dash dotted lines show the contributions in the two-band model of the big gap and small gap, respectively (see text). The blue line corresponds to a single-gap BCS curve. The saturation yields $\lambda _{ab}(0)$ = 445(15)\,nm. Lower panel: The fitting curves (solid lines) were obtained within the following anisotropic $s$-wave, and $d$-wave models of the gap symmetries (see text). The inset presents the temperature dependence of the magnetic penetration depths $\lambda _{ab}(T)$ of FeSe.} } \label{Fig:1}
\end{figure}

\begin{figure}
\includegraphics[width=20pc,clip]{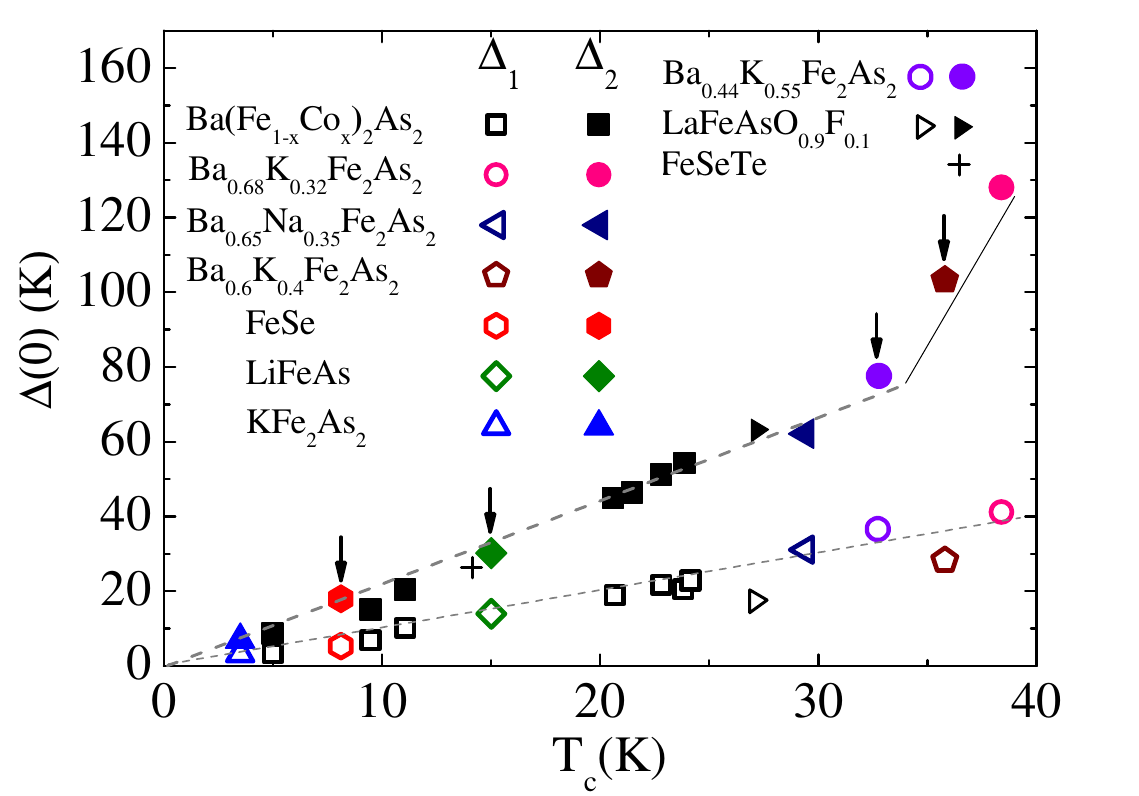}
\caption{The  superconducting-gap evolution of FeSe (this study) together with Ba$_{0.65}$Na$_{0.35}$Fe$_{2}$As$_{2}$,~\cite{Pramanik2011} Ba(Fe$_{1-x}$Co$_{x}$)$_{2}$As$_{2}$~\cite{hardy} for $0.05\leq x \geq0.146$, Ba$_{0.68}$K$_{0.32}$Fe$_{2}$As$_{2}$,~\cite{Popovich2010} Ba$_{0.6}$K$_{0.4}$Fe$_{2}$As$_{2}$,~\cite{CR} KFe$_{2}$As$_{2}$,~\cite{Hafiez2011} LiFeAs,~\cite{YS}, Ba$_{0.45}$K$_{0.55}$Fe$_{2}$As$_{2}$,~\cite{TSh} LaFeAsO$_{0.9}$F$_{0.1}$,~\cite{YaG} and FeSeTe.~\cite{TK} Lines are guides to the eye. The arrows show that both gap values are determined using the lower critical field, $H_{c1}(T)$, studies.} \label{Fig:1}
\end{figure}

Knowing that (i) a single isotropic gap scenario cannot describe our data and (ii) the presence of two superconducting gaps is observed in a variety of different pnictides (See Fig. 5), we applied a phenomenological two-gap model reported by Carrington and Manzano.~\cite{Carrington} Within this model, the temperature dependence of each energy gap can be approximated as:~\cite{Carrington}  $\Delta _{i}(T) = \Delta _{i}(0) {\tanh[1.82(1.018(\frac{T_{ci}}{T}-1))^{0.51}]}$. In the Carrington and Manzano approach the one-band expression is straightforwardly generalized to the two-band case. The obtained experimental temperature dependence of $\lambda^{-2}(T)$ are fitted using a model of two BCS superconducting bands within the clean limit approach for a London superconductor with different gaps.~\cite{Brinkman}. According to Ref.~\onlinecite{VAG}, for each band, $\lambda _{i}^{-2}(T)$  is given by:

\begin{table*}
\caption{\label{tab:table 1} {The superconducting transition temperature $T_{c}$ (K), the $d$-wave approach ($\Delta_{0}$ = meV), the anisotropic $s$-wave approach ($\Delta_{0}$ = meV), two $s$-wave gaps (meV), and the London pentration depth $\lambda _{ab}(0)$ extracted from the the temperature dependence of the London penetration depth for FeSe and FeSe$_{0.85}$.~\cite{RKH}}}
\begin{ruledtabular}
\begin{tabular}{cccccccc}
System &$T_c$ & $d$-wave& anisotropic $s$-wave & two $s$-wave  & $\lambda$(nm) & Ref. \\
\hline
FeSe &  9.4 & 2.045(5)  & 1.663(5) &0.41 and 3.33   &445(15) & This work\\

\hline
FeSe$_{0.85}$ & 8.26 & 1.8(5) & 2.2(3) &0.38 and 1.6  &405(7) &~\cite{RKH}\\

\end{tabular}
\end{ruledtabular}
\end{table*}
\begin{equation}
\label{eq2} \lambda _{i}^{-2}(T) = \frac{\Delta _{i}(T){\tanh(\frac{\Delta _{i}(T)}{2k_{B}T})}}{\lambda _{i}^{2}(0)\Delta _{i}(0)},
\end{equation}
where $\lambda _{i}(0)$ is the residual penetration depth for each band, $k_{B}$ is the Boltzmann constant. Considering different contributions of each band to the whole $\lambda^{-2}(T)$, the following expression was used: $\lambda^{-2}(T) = r\lambda _{1}^{-2}(T) + (1-r)\lambda _{2}^{-2}(T)$ with $r$ being the weighting factor that indicates the contribution of the small gap.

To calculate the theoretical curves, the parameters $\Delta _{1}(0)$, $\Delta _{2}(0)$, and their respective ratios are adjusted. The results of the calculation using the above equations are shown in Fig.\,4. The best description of the experimental data is obtained using values of $\Delta _{1}(0)$ = 0.41 $\pm$ 0.1\,meV, $\Delta _{2}(0)$ = 3.33 $\pm$ 0.25\,meV, $r$ = 0.2. The calculated penetration depth data are represented by the solid red line in the upper panel of Fig.\,4.

{The temperature dependence of the magnetic penetration depth of the anisotropic $s$-wave and $d$-wave gap calculations were performed using the following functional form:~\cite{RKH,RKH2}}

\begin{equation}
\label{eq2} \frac{\lambda _{ab}^{-2}(T)}{\lambda _{ab}^{-2}(0)} = 1 + \frac{1}{\pi}\int_{0}^{2\pi} \int_{\Delta(T,\varphi)}^{\infty} (\frac{\partial f}{\partial E})\frac{EdEd\varphi}{\sqrt{E^{2}-\Delta(T,\varphi)^{2}}},
\end{equation}

\begin{equation}
\label{eq2} f = \frac{1}{[1+\exp(E/k_{B}T)]},
\end{equation}
{where $f$ is the Fermi function, $\varphi$ is the angle a long the fermi surface, and $\Delta(T,\varphi)$ = $\Delta_{0}\delta(T/T_{c})g(\varphi)$ ($\Delta_{0}$ is the maximum gap value at $T$=0). The function $g(\varphi)$ is given by $g^{d}(\varphi)$ = $|cos(2\varphi)|$ for the $d$-wave gap, while $g^{s}(\varphi)$ = (1+$a \cos 4\varphi$)/(1+$a$) for the anisotropic $s$-wave gap.~\cite{Sh1} The results of the analysis are presented in the lower panel of Fig.\,4 by solid lines. The best description of the experimental data for the anisotropic $s$-wave is obtained using values of $\Delta _{0}$ = 1.663(5)\,meV, $a$ = 4.772, and $\lambda _{ab}(0)$ = 430(15)\,nm. For the $d$-wave case, we get $\Delta _{0}$ =  2.045(5)\,meV. It is obvious that the $d$-wave case cannot describe the penetration depth data. On the other hand, the experimental data are well described for both anisotropic $s$- and two-gap $s$ wave models.

It is interesting to compare the extracted values with those obtained previously on the off-stoichiometry compound FeSe$_{0.85}$ (see table 1)}. It is worth mentioning here that the two-gap model describe the in-plane penetration depth data on FeSe$_{0.85}$ with gap values of 1.60 and 0.38\,meV, substantially different to the values we report here for FeSe.~\cite{RKH} {  This might not be surprising since it has been well established that changing of Se content not only leads to a different $T_{c}$ but, as shown by McQueen {\it et al.}~\cite{McQ}, sligth changes from the ideal 1:1 ratio in FeSe, lead to severe changes of the superconducting properties. For instance, the low field magnetization data of various FeSe$_{1\pm\delta}$ samples showed that the strongest superconducting signal occurs for the most stoichiometric sample, whereas it has been shown that for the FeSe$_{0.82}$ case, there is no superconducting signal.~\cite{Will}}

The extracted gap values for the two-gap $s$ wave model are also different from, $\Delta _{1,2}(0)$= 2.5 and 5.1\,meV, reported for FeTe$_{0.55}$Se$_{0.45}$,~\cite{CCH} but comparable with the two-band $s$-wave fit for the multiple Andreev reflections spectroscopy,~\cite{YaG2} ($\Delta _{1,2}(0)$ = 0.8 $\pm$ 0.2 and 2.75 $\pm$ 0.3\,meV). Such a multigap nature seems to be a common scenario for Fe-based superconductors. {It should be noted that both gap values are not far from those reported for LiFeAs single crystals~\cite{YS} but much smaller than those reported in Ba$_{0.6}$K$_{0.4}$Fe$_{2}$As$_{2}$ and Ba$_{0.45}$K$_{0.55}$Fe$_{2}$As$_{2}$.~\cite{CR,TSh}} The lower gap in LiFeAs, Ba$_{0.6}$K$_{0.4}$FeAs, and in our system is smaller but significantly affects the zero-temperature penetration depth. It should be pointed out that such a small gap has also been in line with specific-heat data on a similar FeSe crystal.~\cite{JYLin} The contribution to the in-plane penetration depth data from each band is also shown in Fig.\,4 by dashed and dotted lines, respectively. In fully gapped superconductors, the penetration depth data should show a flat behavior at low temperatures. {However, using the two-band model, we can get an expected saturating behavior below 0.4\,K, indicating a full-gap superconducting state. This saturation yields $\lambda _{ab}(0)$ = 445(15)\,nm, which is somewhat smaller than (560(20)\,nm) in Fe(Te, Se).~\cite{HKi} Our estimated $\lambda _{ab}(0)$ value is indeed comparable with the $\lambda _{ab}(0)$ value (405(7)\,nm) derived from muon-spin rotation studies.~\cite{RKH}} Our results provide another strong evidence that FeSe is not a simple single gap.

Finally, for the sake of comparison the gap amplitudes as a function of $T_{c}$ of FeSe single crystals are shown in Fig.\,5 together with Ba(Fe$_{1-x}$Co$_{x}$)$_{2}$As$_{2}$\cite{hardy} for $0.05 \leq x \geq 0.146$, Ba$_{0.68}$K$_{0.32}$Fe$_{2}$As$_{2}$,\cite{Popovich2010} Ba$_{0.65}$Na$_{0.35}$Fe$_{2}$As$_{2}$,\cite{Pramanik2011}KFe$_{2}$As$_{2}$,\cite{Hafiez2011}LiFeAs,\cite{YS} Ba$_{0.45}$K$_{0.55}$Fe$_{2}$As$_{2}$,\cite{TSh} LaFeAsO$_{0.9}$F$_{0.1}$,~\cite{YaG}and FeSeTe.~\cite{TK} As it can be seen, the gap values differ for different compounds within the 122-family and also for 11, 111, and 1111-compounds. Furthermore, Ponomarev {\it et al}.~\cite{YaG2} have proven that the small and larger gaps increase linearly with $T_{c}$. It is also clear that the larger gap increases stronger than linear with $T_{c}$ for $T_{c}$ $\geq$ 30\,K. In addition, the values of the underdoped, optimally and overdoped K-doped data fit onto the same curves of $\Delta _{1}(0)$, $\Delta _{2}(0)$. On the other hand, for the large gap values, a tendency for strong-coupling effects, e.g., the compounds with the highest $T_{c}$, as in K and Na-doped superconductors is given. Very recently the evolution of the electronic structure of the single-layer FeSe film during the annealing process illustrates that the superconductivity is in the strong-coupling regime. In addition, both the superconducting gap and the transition temperature increase with the annealing process.~\cite{ShHe}{ We do not yet have a good understanding for such large gap behavior especially above 30\,K. However, it is worth mentioning that the contribution of both gap values fit well with the other hole- and electron-doped 122 systems as well as with the 111 and 1111 compounds.}


\subsection{Summary }

In conclusion, we have determined the temperature dependence of the lower critical field $H_{c1}(T)$ of FeSe by the onset of either the trapped moment or nonlinear $M(H)$ response. Assuming either a two $s$-wave-like gaps with magnitudes $\Delta _{1}$ = 0.41 $\pm$ 0.1\,meV and $\Delta _{2}$ =3.33 $\pm$ 0.25\,meV {, or an anisotropic $s$-wave using values of $\Delta _{0}$ = 1.663\,meV ,} we account for the temperature dependence of the lower critical field $H_{c1}(T)$. {These observations clearly show that there is no nodes in the superconducting energy gap of FeSe.} The London penetration depth $\lambda _{ab}(T)$ is calculated from the lower critical field and yields $\lambda _{ab}(0)$ = 445(15)\,nm.


\begin{acknowledgments}
The authors thank Jun Li and Kelly Houben for fruitful discussions. {We thank Paulo de Sousa Pereira for helping us with the calculations}. This work is supported by the FNRS projects, ``cr\'edit de d\'emarrage U.Lg.", the MP1201 COST Action and by the Methusalem Funding of the Flemish Government. The work of A. N. V. and D. A. C. was supported by RFBR grants 13-02-00174 and 12-02-90405. J. V. d. V. acknowledges support from FWO-Vl.
\end{acknowledgments}

\end{document}